\documentclass[12pt]{article}

\usepackage{newtxtext,newtxmath}

\usepackage{graphicx}

\usepackage[letterpaper,margin=1in]{geometry}

\usepackage{xcolor}

\renewcommand{\textcolor}[2]{#2}

\renewenvironment{abstract}
	{\quotation}
	{\endquotation}

\date{}

\makeatletter
\renewcommand{\fnum@figure}{\textbf{Figure \thefigure}}
\renewcommand{\fnum@table}{\textbf{Table \thetable}}
\makeatother

\usepackage{scicite}
\usepackage{caption}
\usepackage{url}

\def\scititle{
	\textcolor{red}{Induction-heated resonant reactors for electrified thermochemistry}
}
\title{\bfseries \boldmath \scititle}

\author{
	Connor~Cremers$^{1\dagger}$,
	Chenghao Wan$^{1\dagger}$,
	Calvin~H.~Lin$^{1}$,
    Zhennan~Ru$^{2}$,\and
    Kesha~Tamakuwala$^{3}$,
    Ariana~B.~Höfelmann$^{1}$,
    Dolly~Mantle$^{4}$, 
    Matthew~W.~Kanan$^{3}$, \and
    Juan~Rivas-Davila$^{1}$, 
    Jonathan~A.~Fan$^{1\ast}$\and
	\small$^{1}$Department of Electrical Engineering, Stanford University, Stanford, CA 94305, USA.\and
	\small$^{2}$Department of Materials Science \& Engineering, Stanford University, Stanford, CA 94305, USA.\and
    \small$^{3}$Department of Chemistry, Stanford University, Stanford, CA 94305, USA.\and
    \small$^{4}$Department of Mechanical Engineering, Stanford University, Stanford, CA 94305, USA.\and
	\small$^\ast$Corresponding author. Email: jonfan@stanford.edu\and
	\small$^\dagger$These authors contributed equally to this work.
}

\begin{document} 
\maketitle

\begin{abstract} \bfseries \boldmath

We present induction-heated resonant reactors, a new concept in electrified thermochemistry in which the reactor itself serves as a volumetric electromagnetic resonator heated through resonant wireless power transfer. \textcolor{red}{We use the Swiss roll resonator as a model system and show that it can be designed to support uniform volumetric heating profiles and enhanced heat transfer characteristics, creating opportunities for process intensification in scaled systems. Compared to conventional (i.e., non-resonant) induction heating systems, resonant reactors can achieve exceptionally high system efficiencies through the combination of near-unity power-to-heat efficiencies and low thermal losses, both enabled by the utilization of resonant energy transfer. These concepts demonstrate how the integration of electromagnetic power transduction with thermochemical reaction engineering enables new opportunities for utilizing green electricity in sustainable chemical conversion.}

\end{abstract}

\noindent
\section*{Introduction}
Thermochemistry is foundational to the modern chemicals industry and is projected to be the dominant mode for chemicals manufacturing for the foreseeable future due to its unparalleled advantages in scaling, cost, and technical maturity \cite{noauthor_thermal_nodate}. The primary means for producing high-grade heat in thermochemical reactors today is the combustion of fossil fuels, which produces carbon dioxide as a byproduct.  To reduce this carbon intensity, an immediate decarbonization strategy is to utilize green electricity for heat \cite{VanGeem2019, Thiel2021}, and electrified heat integration strategies \cite{Mallapragada2023} for reactors based on electrical resistance \cite{wismann_electrified_2019, zheng_joule-heated_2024, Cremers2026Optimal}, thermal plasmas \cite{snoeckx_plasma_2017, wang_plasma_2024}, magnetic induction \cite{truong-phuoc_induction_2025, lin_electrified_2024}, and microwaves \cite{goyal_review_2022, priecel_advantages_2019} have been recent topics of investigation.  Proof-of-concept demonstrations indicate that electrified reactors are capable not just of producing high grade heat to drive chemical reactions, they can also impart new capabilities beyond the limits of conventional reactors such as enhanced heat transfer to catalysts \cite{wismann_electrified_2019, from_electrified_2024} and the driving of reactions under non-equilibrium conditions \cite{nozaki_plasma-enabled_2024,Deng2025}. \textcolor{red}{While much progress has been made, significant limitations remain that hinder the widespread adoption of these concepts. One critical challenge is the limited transduction efficiency of electricity to fields and heat within the reactor, due to inefficiencies in field generation and parasitic heating outside the reactor \cite{Noble2024Radiofrequency}. Another challenge involves the finite penetration of fields into reactor media due to screening and skin depth effects, which lead to non-uniform heating profiles and limitations in reactor scaling \cite{Mallapragada2023, Stankiewicz2020Beyond, Wan2026ScaleUp}.}



\begin{figure} 
	\centering
	\includegraphics[width=\textwidth]{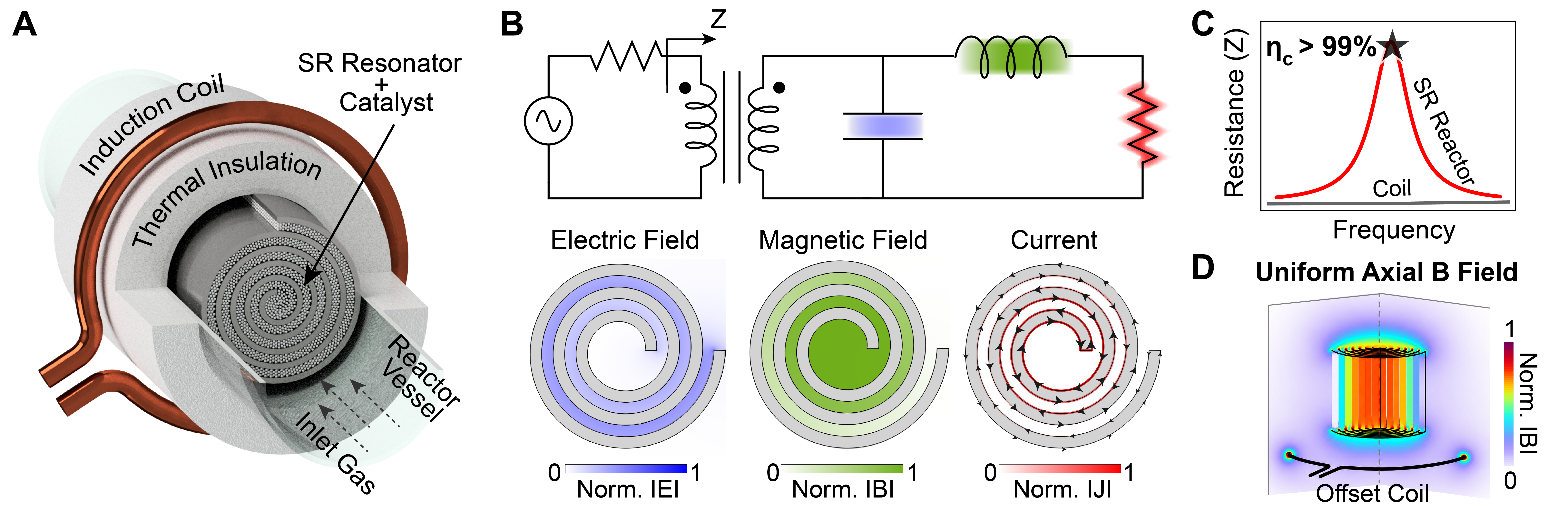} 
	\captionsetup{font=footnotesize}
	\caption{\textbf{Resonant reactors for electrified thermochemistry.}
		(\textbf{A}) Schematic of a resonant flow reactor comprising a Swiss roll (SR) susceptor loaded with catalysts and surrounded by thick thermal insulation. Reactor heating is achieved by resonant wireless power transfer from a single turn induction coil to the Swiss roll.
        (\textbf{B}) Circuit model of the resonant reactor.  The Swiss roll resonator is described as an RLC circuit that supports enhanced energy storage in electric and magnetic fields and the enhanced generation of surface currents, which dissipate as heat in a volumetric profile.
		(\textbf{C}) AC resistances of the coil and Swiss roll resonator, which represent power dissipation in the coil and susceptor, respectively.  With resonant enhancement of the Swiss roll AC resistance, the coupling efficiency is near 100\%.
        \textcolor{red}{(\textbf{D}) Simulated magnetic flux density of a uniform Swiss roll susceptor driven by a single-turn drive coil positioned at its base. The magnetic field distribution within the Swiss roll is volumetric, nearly axially uniform, and is independent of the drive coil position.}
        }
	\label{fig: Fig. 1} 
\end{figure}

We present induction-heated resonant reactors, a conceptually new approach to electrified thermochemistry in which the reactor itself is a volumetric electromagnetic resonator (Fig.~1A). Reactor powering is performed by coupling an external high frequency magnetic field to the reactor via principles in resonant wireless power transfer \cite{Kurs2007}, which induces resonant circulating currents that dissipate as heat.  Power transfer can be described using a circuits picture (Fig.~1B), in which the resonator is described as an RLC loop that supports the resonant amplification of stored electromagnetic energy and the enhancement of induced current amplitude and dissipated power in the resistive load. The relative amount of susceptor-to-coil heating, which we term coupling efficiency ($\eta_c$), is boosted by resonance and is on the order of 99\% (Fig.~1C). Resonant power transfer is widely utilized in the domains of energy harvesting \cite{Rong2025Magnetically, Lu2015Wireless}, wireless charging \cite{Wang2005Design, Dai2017Safe}, and communications \cite{Bi2016Wireless, Degen2021Inductive}, using implementation schemes based on discrete circuit elements, such as inductive loops loaded with discrete capacitors, and with the aim of performing power transfer with minimal heat generation.  In our case, the reactor itself serves as a structural resonator with effective circuit-like properties, and the objective is to dissipate all of the transferred power into high-grade heat.

\textcolor{red}{We demonstrate that resonant reactors provide several unique advantages that, together with near-unity coupling efficiency, address key limitations in electrified thermochemical reactor design. First, the heating profiles of resonant electromagnetic susceptors (i.e., the inductively heated baffle structure in the reaction zone) are fully specified by their electromagnetic mode profiles, independent of the drive coil geometry and position. This enables the detailed volumetric heating profile within the reactor to be tailored by resonator geometry engineering, thereby overcoming the penetration-depth limitations typical of non-resonant energy transfer concepts. Second, the decoupling of susceptor heating profile with drive coil configuration allows for significant system design freedom. For example, the coil can be positioned away from the susceptor to accommodate thick thermal insulation around the reactor, reducing heat losses without affecting the axial uniformity of the magnetic field (Fig. 1D). Third, resonant reactors provide a pathway to scaling, as electromagnetic resonators follow well-defined geometric and frequency scaling rules. Fourth, resonant reactors generate volumetric electric and magnetic fields within the reactor, creating opportunities for field-assisted heating and chemical conversion beyond conventional Joule heating mechanics.}

\section*{Results}
\subsection*{Reactor design methods}
Resonant reactor design broadly encompasses the multiphysics co-design of power electronics, resonant susceptor geometry, heat transfer, and catalyst properties.  Proper selection of the susceptor geometry depends on the chemical conversion process of interest, and we focus this study on the Swiss roll resonant susceptor and its utilization for an endothermic gas reaction performed in a fixed bed reactor. Swiss roll resonators have been a popular topic of study over the last three decades in the electromagnetic metamaterials community, where ensembles of Swiss rolls have been shown to serve as artificial media with geometry-dependent effective magnetic responses \cite{swiss_rolls, Wang2019Broadband, Wiltshire2007Effective}.  Swiss rolls have also been investigated as reactor elements in the chemical engineering community, as they possess large surface areas and resemble a micro-channel sheet reactor rolled up into a cylinder, eliciting predictable laminar fluid flow and exceptional heat transfer coefficients between the walls and channel media \cite{IlKim2007Development, Tseng2023Swiss, Sarkar2011Heat}.  In spite of these research efforts, little work has been done to understand how Swiss rolls dissipate heat when coupled to electromagnetic fields, and no connection has been made in understanding how the wireless heating of Swiss rolls can be harnessed to drive chemical reactions.

\begin{figure} 
	\centering
	\includegraphics[width=\textwidth]{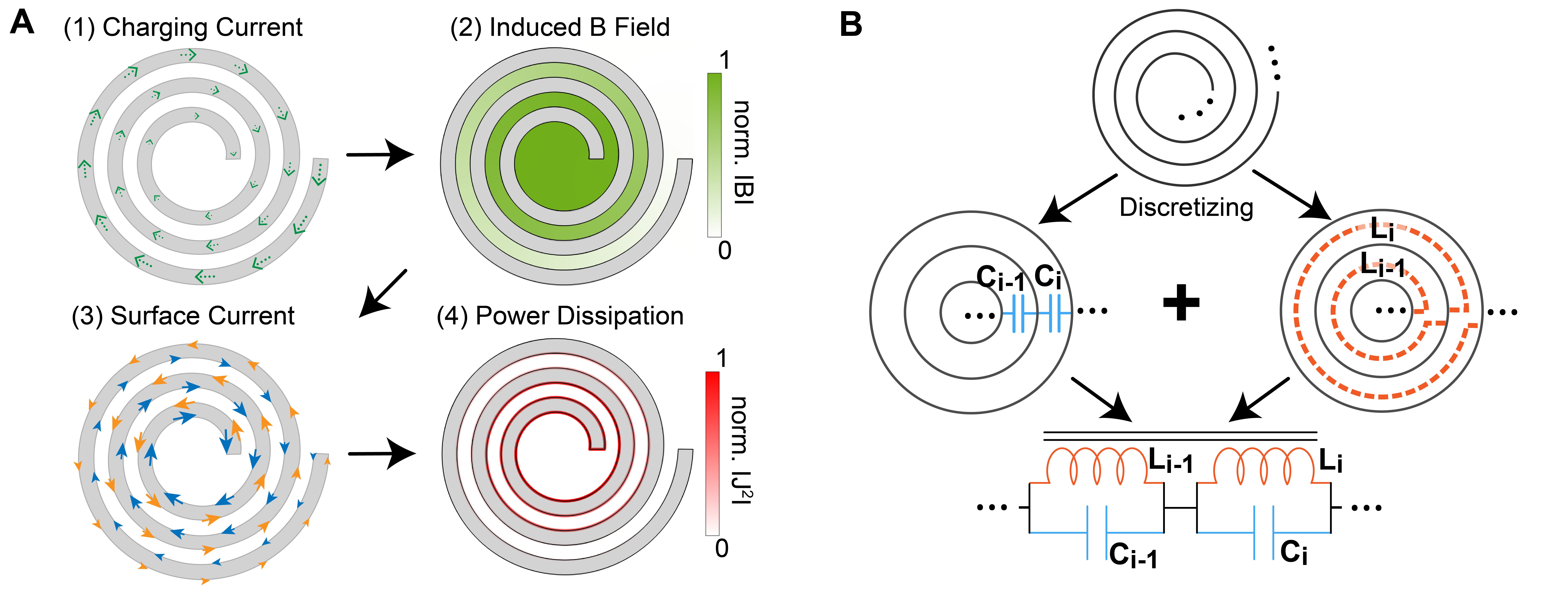} 
	\captionsetup{font=footnotesize}
	\caption{\textbf{\textcolor{red}{Design principles of Swiss roll resonators.}}
		(\textbf{A}) Schematics describing the physics of Swiss roll resonator heating. (1) Coupling of a drive coil to the Swiss roll resonator produces a resonantly enhanced charging current along the metal walls. For a given wall, the charging current represents the sum of currents (i.e., surface currents) along the inner and outer wall surfaces.  
        (2) The magnetic field stored within the Swiss roll is primarily defined by those produced by the charging currents.
        (3) Ampere's law can be applied at the inner and outer surfaces of the resonator walls to calculate the magnitude of surface currents everywhere in the Swiss roll.  
        (4) The power dissipation profile can be readily calculated from the surface current profile using Ohm's law.
        (\textbf{B}) Circuit model of the Swiss roll resonator. The Swiss roll can be modeled as a set of distributed capacitors and inductors that form a series of coupled LC resonators.  The $(i-1)^{th}$ and $i^{th}$ sets of distributed capacitors and inductors are depicted.
        }
	\label{fig: Fig. 2} 
\end{figure}

We first investigate the underlying physics of Swiss roll heating to understand how its volumetric heating profile can be tailored for thermochemical processes. Our effort builds on the Swiss roll resonator model developed by Pendry\cite{swiss_rolls}, which assumes that the inner and outer diameters of the Swiss roll are approximately the same and which does not provide a description of the induced current profiles in the structure.  We begin with a qualitative description of uniform Swiss roll heating at resonance and consider an infinitely long Swiss roll comprising metal with a thickness that is multiple times the skin depth.  In the presence of an external, axially oriented magnetic field, an electromotive force is induced in each turn of the Swiss roll, which drives currents in the structure via Faraday induction (Fig. 2A, (1)). We term these currents as ``charging currents," and they account for the combined currents flowing along both the inner and outer metal surfaces of each turn.  The outer turns feature larger cross sectional areas and therefore larger electromotive forces compared to the inner turns, leading to enhanced charging currents in the outer turns.

The magnetic field distribution within the Swiss roll is dominated by the fields induced by the charging currents (Fig. 2A, (2)), which have substantially larger magnitudes compared to the external magnetic field due to resonant enhancement from the Swiss roll.  The total induced magnetic fields are the superposition of fields produced by each turn, and as such, the induced magnetic field for a uniform Swiss roll is largest in the center of the roll. These induced field values, together with Ampere's law, can be used to quantify the amplitude of currents at the inner and outer metal surfaces of each turn (Fig. 2A, (3)).  We term these interfacial currents as ``surface currents," and their dissipation as heat determines the volumetric heating profile of the Swiss roll (Fig. 2A, (4)).  The surface and charging current pictures are self-consistent, as the sum of surface currents along the inner and outer surfaces of the metal equals the charging currents.  Based on this understanding, we find that uniform Swiss rolls produce a centrally heated volumetric profile, which is counter to the seemingly intuitive picture of an outer edge heating profile specified by the charging currents.

\begin{figure} 
	\centering
	\includegraphics[width=\textwidth]{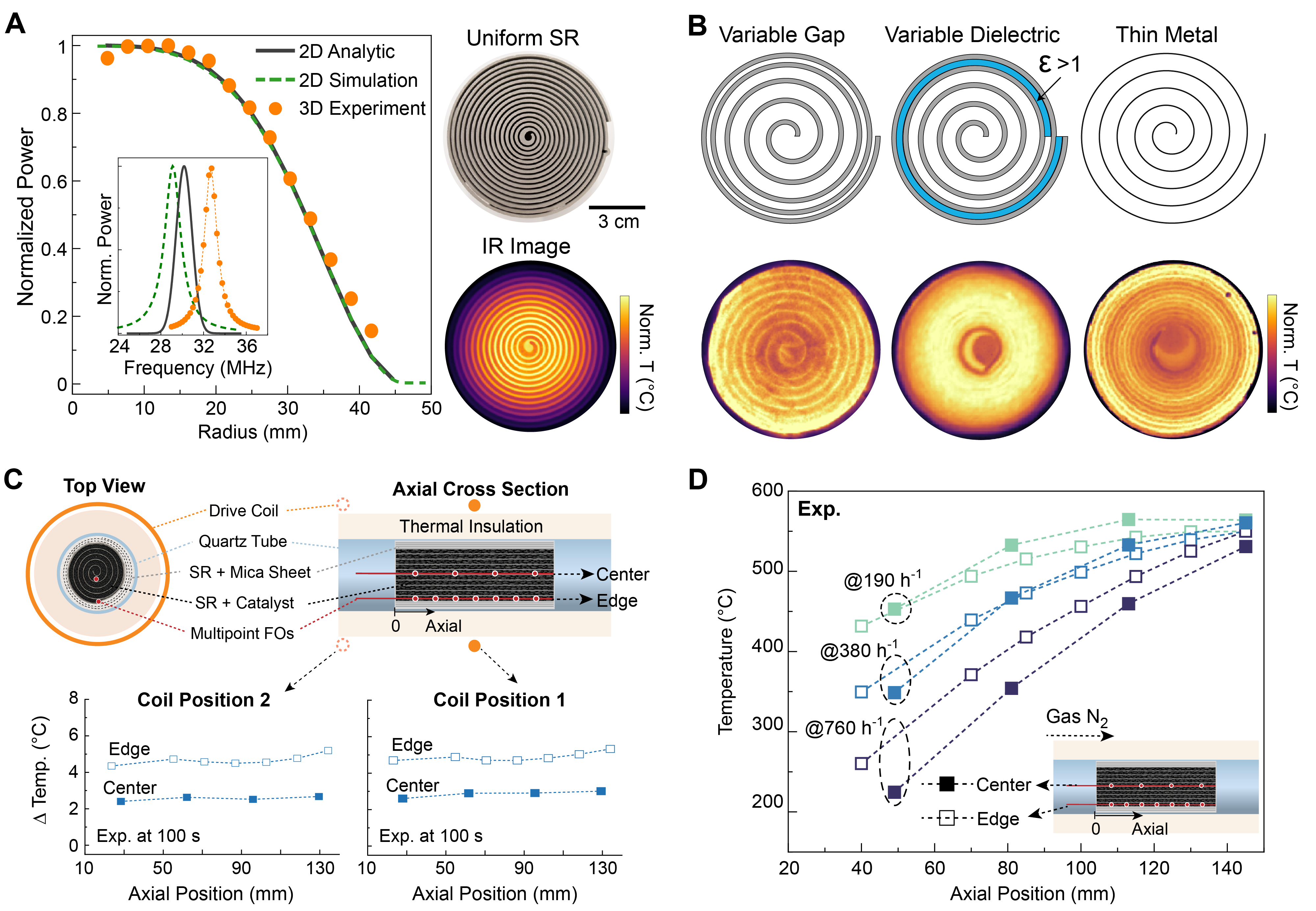} 
	\captionsetup{font=footnotesize}
	\caption{\textbf{\textcolor{red}{Heating characteristics of uniform and inverse designed Swiss roll resonators.}}
		(\textbf{A}) Theoretical and experimental heating profile of the uniform Swiss roll.  Analytical, numerical, and experimental temperature profiles are in agreement and show a centrally heated profile. Total power dissipation as a function of frequency (inset) shows that the 2D analytic and numerical models have consistent resonance frequencies and quality factors.  The frequency shift displayed by the 3D resonator spectrum is due to fringe fields not present in the 2D models.
        (\textbf{B}) 
        Schematics and experimental temperature images of Swiss roll resonators exhibiting uniform volumetric heating profiles.  Resonators are inverse designed using variable gap widths (left), variable dielectric loading (center), and thin metal walls with thickness less than the metal skin depth (right).
        \textcolor{red}{(\textbf{C})  Schematic (top) of the uniformly heated Swiss roll reactor, based on variable mica loading, with axially-aligned multi-point fiber optic temperature probes along two radial positions the resonator.  Plots of transient temperature measurements (bottom) are taken with the magnetic drive coil placed at different axial positions.}
        \textcolor{red}{(\textbf{D}) Steady state volumetric temperature profile of the Swiss roll resonator heated under different nitrogen flow rates.  The outlet temperature is set to 550 $^\circ$C.}
        }
	\label{fig: Fig. 3} 
\end{figure}

A quantitative framework for Swiss roll physics can be constructed by modeling the resonator as a set of concentric rings, which can be described as a distributed circuit comprising a series of coupled LC resonators (Fig. 2B). The distributed inductance derives from the geometry of individual Swiss roll turns, while the distributed capacitance derives from the geometry of pairs of adjacent turns. The voltage load across each LC resonator can be determined using Faraday's law.  All of these terms can be approximately expressed in simple analytic forms, and the resulting circuit model can be used to solve for the charging current profile and resonance frequency. The induced magnetic fields everywhere and the surface currents and ohmic losses in the Swiss roll metal layers are subsequently calculated using Maxwell's equations.  The quality factor of the Swiss roll resonator, which correlates with coupling efficiency, can be directly calculated as field energy stored in the reactive circuit elements divided by the energy dissipated per cycle due to surface current heating. Importantly, our Swiss roll circuit description can generalize to the modeling of asymmetric resonators that feature radially dependent gap widths and dielectric loading profiles. 
Utilization of this formalism in an inverse design framework enables customized radial heating profiles to be specified using intentional symmetry breaking.  

\textcolor{red}{To validate our understanding of Swiss roll heating and our analytic circuit model, we experimentally investigate the heating profile of Swiss roll resonators fabricated out of stainless steel.  We first image the surface heating profile of the Swiss roll with uniform turn spacing using transient thermal imaging, in which thermal imaging is performed after only one minute of heating to prevent heat transfer processes from distorting the imaged profile.  The uniformly-spaced Swiss roll displays a center-heated profile, which quantitatively matches with our analytic circuit model and numerical simulations (Fig. 3A). We then utilize our analytic circuit model to perform inverse design and construct Swiss roll configurations that are explicitly designed to support uniform heating profiles. Such a configuration pushes the process intensification limits of resonant reactors by minimizing radial heating gradients in the system.  The uniformly heated Swiss rolls are numerically designed using our circuit model, and the resulting analysis indicates that uniform heating can be achieved when the distributed capacitance at the outer turns is enhanced. We consider methods for capacitance enhancement based on variable gap width tuning and dielectric loading with mica. We also find with our theory that uniform heating can be achieved with uniform Swiss rolls when the metal thickness is much less than the skin depth.  Transient infrared images of the three types of uniformly heated Swiss role structures are presented in Fig. 3B and all show heating profiles with enhanced uniformity.}  

\textcolor{red}{We further probe the volumetric heating profile within the uniformly heated Swiss roll resonator, based on selective dielectric loading with mica and otherwise loaded with alumina pellets mimicking catalyst particles. 
We first apply our transient temperature measurement scheme with multi-point fiber optic temperature sensors placed near the axial center and edge of the resonator.  The resulting temperature profiles, measured 100 seconds after the initiation of heating, are shown in Fig. 3C and feature a high degree of axial uniformity.  The temperature magnitudes along the axial center and edge are comparable, indicating a relatively high degree of radial heating uniformity. Differences in the transient temperature magnitude at these two axial positions are attributed to nonidealities in experimental Swiss roll manufacturing and heat capacity differences in the Swiss roll environment at these positions.
We additionally take these measurements with the magnetic drive coil placed at two different axial positions and find that the susceptor heating profile is insensitive to coil position, confirming that the heating profile is decoupled from the magnetic coil configuration.}

\textcolor{red}{We probe the impact of our uniform volumetric heating capability on the volumetric susceptor temperature profile at reaction-relevant conditions by heating the reactor to 550 $^\circ$C while flowing nitrogen gas.  The results are shown in Fig. 3D for different gas flow rates and demonstrate that the temperature profile is nearly radially uniform in all cases, indicating that the high degree of heating uniformity supported by the Swiss roll resonator enables nearly ideal temperature profiles within the reactor even under gas flow conditions. We note that the heating profile can undergo minor changes with increasing temperature due to the temperature dependence of the metal resistance and catalyst dielectric properties.}

\begin{figure} 
	\centering
	\includegraphics[width=\textwidth]{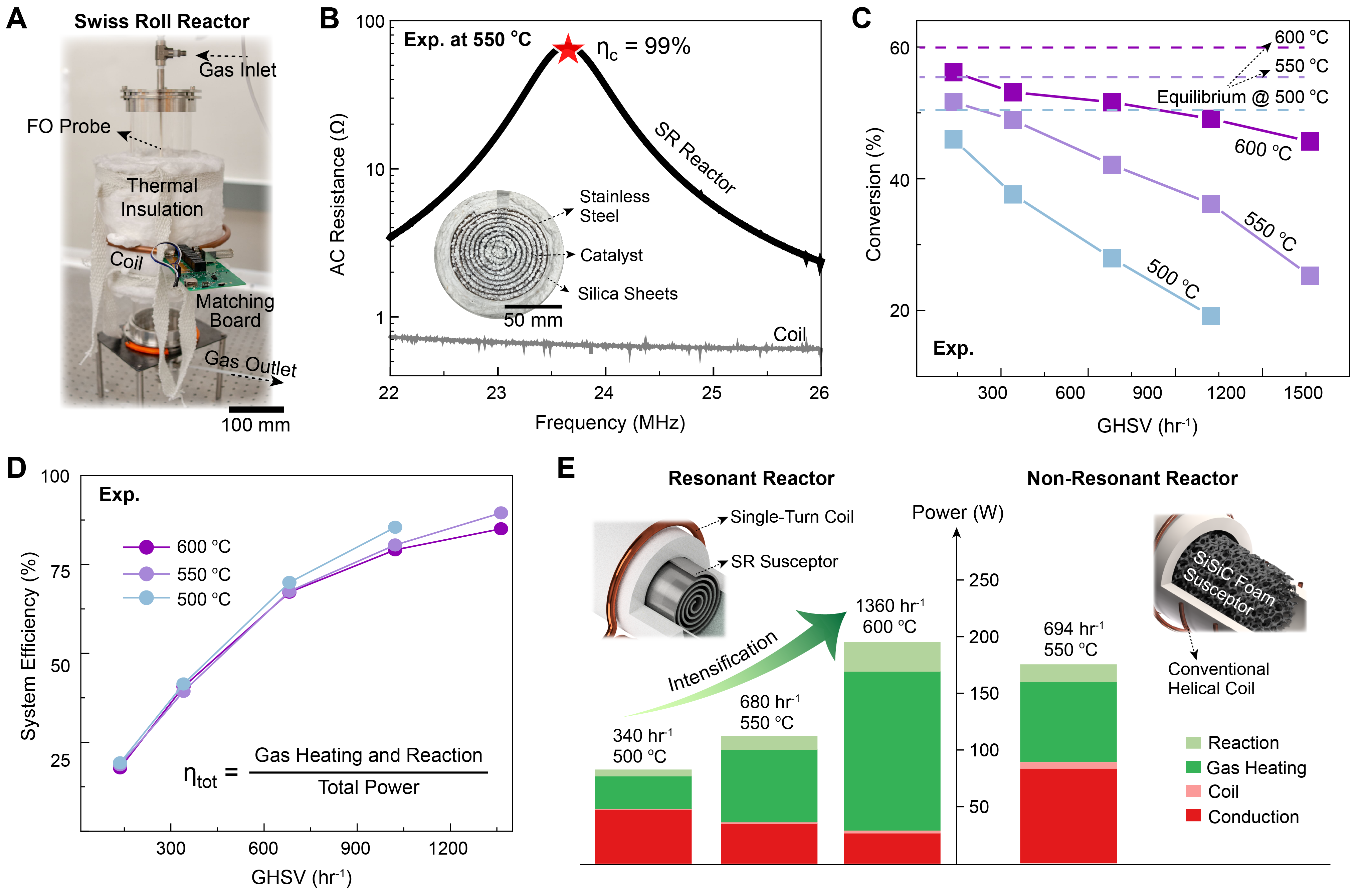} 
	\captionsetup{font=footnotesize}
	\caption{\textbf{\textcolor{red}{Experimental Swiss roll RWGS flow reactor demonstration.}} 
        (\textbf{A}) Experimental image of the packed Swiss roll flow reactor. The Swiss roll comprises 13 turns and is tailored using variable gap widths to support a uniform volumetric heating profile.  The resonator is loaded with 1 mm-diameter catalyst particles.  The reaction zone is surrounded by 55 mm-thick thermal insulation, followed by a single turn magnetic drive coil.
		(\textbf{B}) Experimental AC resistance measurement of the Swiss roll reactor and drive coil as a function of frequency, for the reactor heated to 550 $^\circ$C.  At the resonance peak, the coupling efficiency ($\eta_c$) is measured to be 99\%.
        Inset: experimental image of the packed Swiss roll susceptor. 
        (\textbf{C}) Experimental CO$_2$ conversion to CO as a function of inlet gas flow for three different outlet temperatures.  The system was run for 1 hour at each condition. For low flow rates, conversion is near thermal equilibrium values. 
        (\textbf{D}) Experimental total system efficiency ($\eta_{tot}$) as a function of flow rate for three different outlet temperatures.  $\eta_{tot}$ is defined as the energy consumed to heat the gas and perform the reaction divided by the energy inputted into the drive coil.  $\eta_{tot}$ experimentally approaches 90\% for high gas flow rates.  
        (\textbf{E}) \textcolor{red}{Performance comparison between the Swiss roll resonant reactor (this work) and a representative non-resonant induction-heated metamaterial reactor\cite{lin_electrified_2024, Wan2026ScaleUp}. 
        For reactors utilizing comparable reaction conditions (dashed boxes: 550\,${}^{\circ}$C and ~680 hr$^{-1}$ GHSV operating conditions), the total system efficiency increases from 45\% in the non-resonant reactor to 67\% in the resonant reactor, due to the combination of reduced coil losses and increased thermal insulation.  
        Light green: energy consumed by the reaction. Dark green: energy consumed by gas heating. Orange: heat dissipation in drive coil. Red: heat loss through the reactor walls.}	
        }
        
	\label{fig: Fig. 4} 
\end{figure}

\subsection*{Swiss roll reactor implementation and reaction results}
With our established understanding of Swiss roll susceptors, we experimentally implement a uniformly heated Swiss roll reactor configured for the reverse water gas shift (RWGS) reaction.  The RWGS reaction utilizes carbon dioxide to produce carbon monoxide, and it has attracted much attention due to its potential role in a circular carbon manufacturing economy \cite{GonzalezCastano2021Reverse}.  We construct and place a 10 cm-diameter, 10 cm-tall stainless steel Swiss roll susceptor in a quartz tube, and the Swiss roll is specified to support uniform volumetric heating via variable gap width tuning.  Fixed bed catalyst particles are loaded into the inner region of the Swiss roll (Fig. 3A), which contains constant 4 mm-wide gap widths.  In this reaction zone, the surface area of metal is 2.65~m$^2$ per meter, which ensures a low heat transfer resistance. The catalyst comprises potassium carbonate supported on mesoporous alumina, which was recently discovered to serve as an effective RWGS catalyst that suppresses methanation and coking side reactions \cite{Li2022Carbonate, Tamakuwala2025Intermediate}.  The quartz tube is wrapped with 70 mm-thick ceramic fiber thermal insulation, followed by a single turn 250 mm-diameter copper drive coil connected to a high frequency power amplifier.  

The experimental heating, energy transfer, and reaction properties of the resonant reactor match well with theoretical prediction.  
AC impedance measurements of the coil and Swiss roll, taken at 550\,${}^{\circ}$C, show a resonance frequency of $23.65$~MHz, which is close to our analytic model prediction of $24.7$~MHz (Fig. 4B). At resonance, the real impedance of the Swiss roll is $67.6~\Omega$ and the drive coil resistance is $0.65~\Omega$, yielding a coupling efficiency of approximately $99\%$.  We perform the RWGS reaction by flowing a 3:1 ratio of H$_2$ to CO$_2$ with $20\%$ Ar as an analysis gas, and closed loop control sets the reactor outlet temperature to values between 500\,${}^{\circ}$C and 600\,${}^{\circ}$C. Chemical conversion for different flow rates and exit temperatures, plotted in Fig. 4C, approaches equilibrium values with low flow rates and drops with increasing flow. 

\textcolor{red}{An analysis of the total system efficiency ($\eta_{tot}$) is presented in Fig. 4D and shows that our resonant reactor supports exceptional efficiency metrics.  In evaluating system efficiency, sensible heating of the inlet gases and the heat of reaction is considered useful, while heat dissipated in the drive coil and heat lost from the reactor by heat conduction through its walls is considered waste.} We observe that at high flow rates, our system efficiency reaches almost $90$\%, which is exceptionally high for a lab scale thermochemical system. In this limit, power losses in the drive coil constitute approximately 1\% of total energy consumption, and approximately 10\% of the power is lost by heat conduction through the thermal insulation. Higher efficiencies can be readily achieved by using even thicker thermal insulation.
\textcolor{red}{To benchmark this result, we compare it with the experimental system efficiencies of a previously reported non-resonant induction-heated metamaterial reactor, which consists of a silicon carbide lattice baffle that is volumetrically heated with a helical magnetic coil (Fig.~4E) \cite{lin_electrified_2024, Wan2026ScaleUp}.
With non-resonant induction heating, efficient coil coupling to the susceptor requires the coil to closely conform to the susceptor geometry, 
necessitating a trade-off between coupling efficiency and thermal heat loss management. 
A direct comparison of representative resonant and non-resonant reactor configurations driving the RWGS reaction under similar conditions (Fig.~4E, dashed boxes) indicates the resonant reactor achieves an $\eta_{\mathrm{tot}}$ of approximately 67\%, compared to 45\% for the non-resonant system, with enhanced energy efficiency a result of both reduced coil losses and improved thermal management.
}

\section*{\textcolor{red}{Resonant reactor scaling}}
We present Swiss roll resonator scaling concepts that represent two limits in scaling methodology. The first involves increasing the number of turns in the Swiss roll while fixing the gap width between the turns.  This method has the advantage of maintaining a large and nearly constant susceptor surface area-to-reactor volume ratio, enforcing constant and extreme microreactor-like heat transfer properties in scaled systems.  Simulated plots of the resonance frequency and quality factor of a Swiss roll susceptor with a constant 4 mm-wide gap are shown as a function of reactor radius in the top panel of Fig. 5A. The resonance frequency decreases roughly as the square of the number of turns, which is advantageous from the viewpoint of power electronics where relatively high power, high efficiency amplifiers can be implemented at kilohertz frequencies \cite{Wang2020Review}.  \textcolor{red}{The decrease in quality factor with scale is proportional to $g\sqrt{\sigma f}$ where $g, \sigma,$ and $f$ are the gap width between turns, electrical conductivity, and frequency, respectively. This trend leads to a corresponding decrease in coupling efficiency, setting practical size limits for this reactor scaling concept.}

\begin{figure} 
	\centering
	\includegraphics[width=\textwidth]{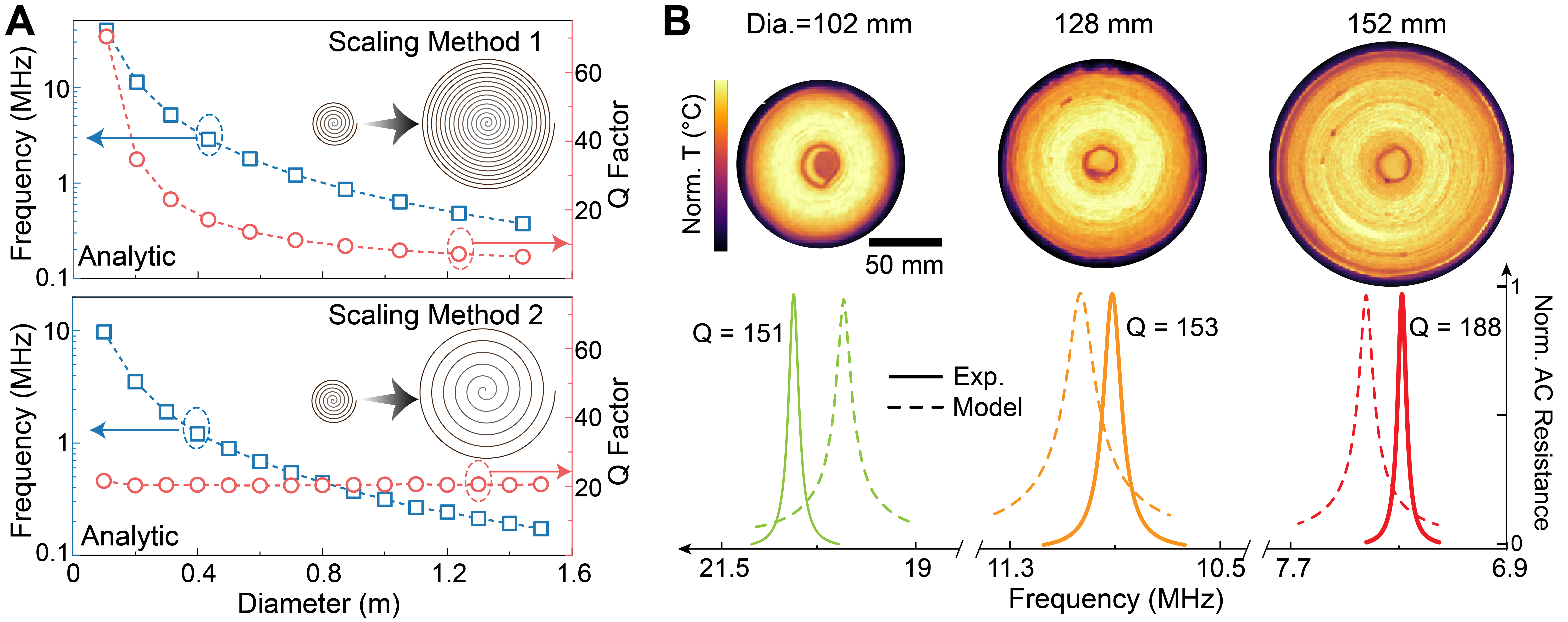} 
	\captionsetup{font=footnotesize}
	\caption{\textbf{\textcolor{red}{Scale up analysis.}}
		(\textbf{A}) Calculated Swiss roll resonance frequency and quality factor as a function of resonator diameter, for two different scaling methods. For scaling method 1 (top), the gap width is fixed to be 4 mm, and for scaling method 2 (bottom), the gap width scales linearly with resonator diameter.
        \textcolor{red}{(\textbf{B}) Experimental (bold) and calculated (dashed) normalized AC resistance and heating profile images for three different Swiss rolls, each with different diameters and each designed to support uniform volumetric heating. For this demonstration, the metal thickness for all Swiss rolls is constant and on the order of the skin depth, leading to an increase in resonator quality factor as a function of scale.} 	
        }
	\label{fig: Fig. 5} 
\end{figure}

In our second method, we scale the gap width and metal thickness linearly.  Simulated plots of the frequency and quality factor versus reactor radius (bottom panel of Fig. 5A) show that the quality factor is independent of scale, enabling a Swiss roll scaling trend that does not suffer from a penalty in coupling efficiency and can therefore extend to very large sizes.  The decrease in resonance frequency with scale indicates that these scaled reactors benefit from the use of highly efficient, kilohertz frequency power electronics, similar to the case of the first scaling method. We note that in comparison to the first scaling strategy, this second method has clear advantages with regards to coupling efficiency, but it also features a reduction in heat transfer coefficient that decreases inversely with reactor diameter. 

We experimentally investigate the scalability of Swiss roll resonators by fabricating and characterizing three rolls of increasing size (Fig. 5B). Resonator geometry scaling follows the first scaling method, and we construct the Swiss rolls using thin metal (50-$\mu$m aluminum foil) spaced by a 3 mm-thick polyethylene foam for ease of manufacturing.  Uniform heating is achieved using a combination of variable dielectric loading, which is achieved by loading the outer two turns of the Swiss roll with mica, and by the use of metal walls that are on the order of or thinner than the skin depth.
\textcolor{red}{We note that as the roll grows larger and the frequency drops, the skin depth increases relative to the metal thickness, which is constant for the different Swiss rolls, leading to reduced power dissipation and an increase in quality factor.  This trend is distinct from that in Fig. 5A, which assumes the metal thickness is always much greater than the skin depth. Both the resonance frequency and quality factor trends align well with our analytic model (dashed lines in Fig. 5B). Transient thermal imaging confirms uniform heating for each fabricated roll.}

\section*{\textcolor{red}{Resonant reactors as field reactors}}
\textcolor{red}{Lastly, we discuss the potential for resonant reactors to generate volumetric electric and magnetic fields for use in field-enhanced chemistries.
Heating concepts based on the coupling of electric, magnetic, and electromagnetic fields to catalysts and reactants have attracted significant interest, due to their potential to selectively heat catalysts and reactants with high power densities \cite{goyal_review_2022, Wang2019Induction, Kuhwald2022Inductive}.  There also exist potential non-thermal enhancements in catalytic kinetics enabled by electromagnetic fields \cite{Dudley2015Microwave, Turnhout2025Plasma}. 
However, widespread implementation of such concepts remains challenging because it remains difficult to generate  large magnitude electric and magnetic fields with volumetric uniformity and high system efficiencies.}

\textcolor{red}{Swiss roll resonators offer a unique means of engineering the volumetric electric and magnetic field distributions within the reaction zone. As a resonator, the field magnitudes get amplified proportionally to the quality factor of the resonator, enabling large field amplitudes to be produced with modest power electronic inputs.
A nearly uniform magnetic field can be generated by concentrating the charging currents into the outer turns of the resonator through either selective dielectric loading or tightly spaced outer turns (i.e., high capacitance between the turns), as illustrated in Fig.~6A. Since each charging current loop amplifies the magnetic field within its interior, this configuration produces an approximately uniform magnetic field throughout the reactor volume. A nearly uniform electric field can be achieved by introducing a high-permeability core that equalizes the magnetic flux passing through each turn of the resonator (Fig.~6B), resulting in a nearly uniform voltage drop and electric field distribution between adjacent turns.} 

\textcolor{red}{When resonant reactors are loaded with electromagnetically lossy media, energy dissipation in the system gets split between the selective heating of the lossy media and Joule heating of the resonator walls.  We analyze the nature of this split by considering field reactors loaded with effectively homogeneous dielectric media with a range of material loss values.  In the limit where the Swiss roll resonators are infinitely long (i.e., fringe fields can be neglected) and there is full overlap between the lossy media and fields within the resonator, the relative heating of the lossy media versus the resonator walls is the same for both electric and magnetic field reactor configurations and depends on only two factors: the media loss tangent and the resonator quality factor.  Relative energy dissipation in the lossy media and resonator walls as a function of loss tangent in this limit is plotted in Fig. 6C for a resonator with a quality factor of 100, which is consistent with our experimental Swiss roll structures.  The plot indicates that when the loss tangent is 0.1 and greater, over 90\% of the input power is dissipated in the lossy medium.  This loss tangent threshold can be adjusted by modifying the resonator quality factor and relative fill fraction of media in the resonator system.  Future research will investigate in more detail how field-based resonant reactors can be implemented in practice.}

\begin{figure} 
	\centering
	\includegraphics[width=\textwidth]{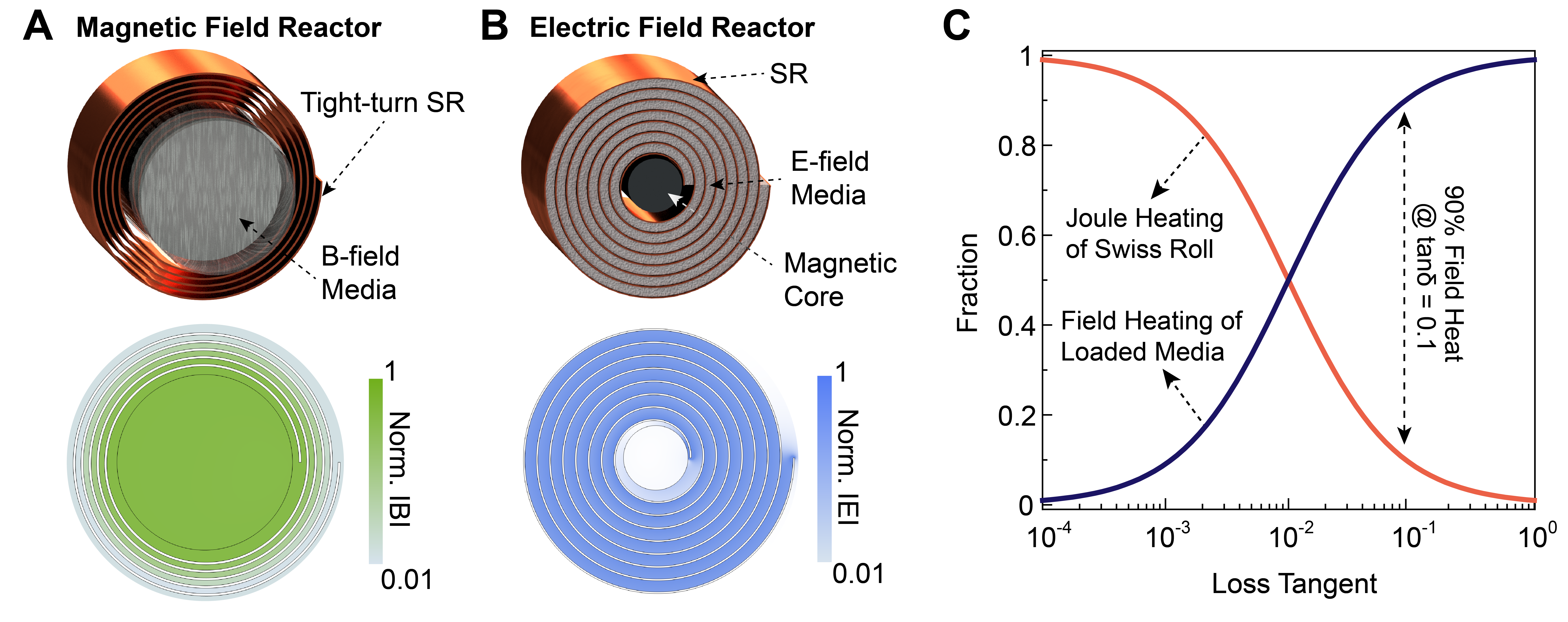} 
	\captionsetup{font=footnotesize}
	\caption{\textbf{\textcolor{red}{Field-assisted reactor configurations.}}		
        \textcolor{red}{(\textbf{A}) Magnetic field Swiss-roll reactor design featuring a uniform, amplified magnetic field within the reaction zone, produced by the resonant charging currents in the tightly-spaced outer turns of the copper sheets.}
        \textcolor{red}{(\textbf{B}) Electric field Swiss-roll reactor design featuring a uniform, amplified electric field between adjacent turns, enhanced by introducing a high-permeability magnetic core that equalizes the magnetic flux throughout the resonator.}
        \textcolor{red}{(\textbf{C}) Analysis of relative energy dissipation in the Swiss roll susceptor and in lossy media loaded in the resonator, as a function of loaded media loss tangent. The media is assumed to fully pack the resonator and a resonator quality factor of 100 is assumed.  Over 90\% of the input energy dissipates in the lossy media as heat when the loss tangent is greater than 0.1.}        
        }
	\label{fig: Fig. 6} 
\end{figure}

\section*{\textcolor{red}{Conclusion}}
We present the concept of resonant reactors, which pushes electrified thermochemistry towards ideal limits in heating, heat transfer, and energy transduction efficiency.  We use Swiss roll resonant reactors as a model system and show that they exhibit multiple desirable characteristics, including 99\% conversion of electricity to internal heat, a large surface area that supports microreactor-like heat transfer properties\cite{Natarajan2019Overview, Wei2025Review}, a relatively low resonance frequency for its size, largely contained and uniform electric and magnetic fields, and scalable manufacturability. Importantly, these structures can support tailored volumetric heating profiles, and we utilize an analytic circuits-based inverse design approach to experimentally demonstrate resonant reactors featuring uniform volumetric heating profiles and exceptional system efficiencies for endothermic gas-phase reactions.  A scale up analysis indicates that our concepts can scale and drive highly endothermic reactions while minimizing radial temperature gradients, which is only possible in a fixed bed system by explicitly defining the volumetric heating profile of the reactor to be uniform and by eliminating heat transfer bottlenecks.

We envision many significant opportunities in resonant reactor design and utilization moving forward. On the reactor design side, there are many types of electromagnetic resonator geometries that can be co-optimized with endothermic chemical conversion processes, which span a wide range of  multiphase flow regimes and reactor configurations. Resonant reactors can also be configured to support multifunctional and reconfigurable capabilities, by designing and driving resonant susceptors with multiple electromagnetic modes and by actively reconfiguring the resonators by mechanical or electrical tuning. Configuration of the resonant reactor as a field reactor presents ample opportunities to deploy field-enhanced chemistries in a scalable manner, where the specific coupling of megahertz frequency electric and magnetic fields to catalysts and reactants present particular opportunities in high efficiency hysteresis and dielectric heating. We additionally see opportunities to adapt our proposed concepts of structural resonant power transfer to other domains of clean energy and manufacturing, where the selective and efficient heating of volumetric structures has utility in electrochemistry, materials synthesis, thermal storage, and other fields.

\clearpage 

%
\bibliography{references} 
\bibliographystyle{sciencemag}

%
%
%
%
%
%


\section*{Acknowledgments}

\paragraph*{Funding:}
Department of Energy, award number: DE-EE0011191 (JAF, MWK, JR)\\
EPIXC, agreement number M2505270 (JAF)\\
Shell Global Solutions Inc., agreement number CW783871 (JAF)\\
Stanford School of Sustainability Accelerator 266899 (JAF) \\
Stanford School of Sustainability Accelerator 266972 (MWK)\\
National Science Foundation ECCS-2026822 (Stanford Nano Shared Facilities)\\
Gates Foundation Fellowship (CHL) \\
Stanford Graduate Fellowship (CC, AH) \\
TomKat Center for Sustainability Graduate Fellowship (DM) \\
\paragraph*{Author contributions:}
Conceptualization: JAF, CC, CW \\
Reactor setup: CC, CW, CHL, ZR, DM, ABH \\
Reactor modeling: CC, CW \\ 
Power electronics: CC, CHL \\
Catalyst preparation: KT \\
Supervision: JAF, JR, MWK \\
Writing: JAF, CW, CC, CHL, ZR, DM, ABH, KT, MWK, JR \\


\end{document}